\documentclass[prl]{revtex4}%
\usepackage{epsfig}
\usepackage{graphics}
\begin{document}
\title{Nuclear Astrophysics of Worlds in the String Landscape}
\author{Craig J. Hogan}
\affiliation{Astronomy and Physics Departments, 
University of Washington,
Seattle, Washington 98195-1580}
\begin{abstract}
Motivated by    landscape  models  in string theory, cosmic nuclear evolution is analyzed allowing the Standard Model Higgs expectation value $w$ to take values different from that in our world ($w\equiv 1$), while holding the Yukawa couplings fixed.  Thresholds are estimated, and astrophysical consequences are described, for several sensitive dependences of nuclear behavior on $w$.  The dependence of the neutron-proton mass difference on $w$ is estimated based on recent calculations of strong isospin symmetry breaking, and is used to derive the threshold of neutron-stable worlds, $w\approx 0.6\pm 0.2$.  The effect of a stable neutron on nuclear evolution in the Big Bang and stars is shown to lead to radical differences from our world, such as a predominance of heavy $r$-process and $s$-process nuclei and a lack of normal galaxies, stars and planets.       Rough estimates are reviewed of  $w$ thresholds for deuteron stability and the  $pp$ and $pep$ reactions dominant in many stars. 
A simple model of nuclear resonances is used to estimate the $w$ dependence of  overall carbon and oxygen production  during normal stellar nucleosynthesis; carbon production is estimated  to change by a fraction $\approx 15(1-w)$. Radical changes in astrophysical behavior seem to require changes in $w$ of more than a few percent, even for the most sensitive phenomena.
\end{abstract}
\maketitle
\section{introduction}

Advances in string theory and cosmology have recently inspired increasing open-mindedness about what is reasonable to expect from a fundamental theory of nature; indeed, it has even been suggested\cite{Weinberg:2005fh} that  ``we may be at a new turning point, a radical change in what we accept as legitimate foundation for a physical theory.''.  The  idea gaining acceptance\cite{Wilczek:2005aj} is that some seemingly fundamental parameters of nature may never be predicted from first principles,  but are determined by selection from an ensemble of universes.
 Part of this program involves studying the astrophysics of counterfactual worlds. This paper studies  some particularly sudden   changes in nuclear astrophysics that occur if the expectation value of the Higgs field, and nothing else in the Standard Model, has a value different from that in our world.

The complete Standard Model with the recently added neutrino degrees of freedom has about twenty-six numerical  parameters that are not currently derived from any fundamental symmetry, but are simply set to fit experimental data.   With cosmological parameters added for quantities such as cosmological constant, dark matter and fluctuation amplitude, the  entire world model is specified by about thirty-one numbers.\cite{Wilczek:2004cr,Tegmark:2005dy} 
Remarkably however,  almost all  behavior of matter important to everyday life---  common nuclear, atomic and molecular physics,   chemistry and biology---  is shaped by just the gauge couplings and  three light fermion masses:  those of the electron, and the up and down quarks.
\cite{Cahn:1996ag,Hogan:1999wh}   
The behavior  of our world depends sensitively on the values of these masses, through the delicately balanced mass differences in the system of electrons, neutrons, protons,  deuterons, and heavier nuclei. In the Standard Model, these three numbers are determined by four parameters. The masses are all proportional to  $w$, defined here as the ratio of  the Higgs expectation value to the QCD scale, 
in units where our universe has $w=1$; and they are independently proportional to  three Yukawa couplings, denoted here by the mass values in our $w=1$ world, $m_e^1$, $m_u^1$, $m_d^1$, as defined in more detail below.

Some of the standard model parameters may turn out to be derivable, especially the dimensionless couplings; it is widely believed that these may arise from symmetry breaking and  renormalization within the context of a unified theory. \cite{Dimopoulos:1981yj}  
 However, a   leading hypothesis for setting dimensionful parameters, such as $w$ (or the cosmological constant\cite{Weinberg:1987dv})  is that their values are determined by selection in a string-landscape multiverse.\cite{Weinberg:2005fh,Wilczek:2005aj,Bousso:2000xa,Giddings:2001yu,Kachru:2003aw,Banks:2003es,Susskind:2003kw,Dienes:2004pi,Douglas:2003um,Ashok:2003gk,Dvali:2003br,Denef:2004ze,Arkani-Hamed:2005yv}\footnote{The $w$ used here is dimensionless and  not dimensionful, but that is  because for the current arguments we are choosing for convenience to define it in units  of the strong interaction scale. That scale, where the strong interactions become strong, is itself derivable from some intrinsic coupling constant at high energy, plus renormalization or ``running''. In the friendly landscape context, this scale is thus determined by the coupling constant which does not scan.} String theory admits multitudinous solutions for stable vacua, constituting a vast landscape of possible universes with different possible couplings, symmetries, and spacetime configurations.  The range of choice may be simplified if we live in a ``friendly'' region of the landscape, over which many string vacua are realized, which share similarly structured  Lagrangians (within the region) but   densely populate  an ensemble  of different potentials.\cite{Arkani-Hamed:2005yv} From the point of view of these models, $w$ is then a random variable ``scanned'' in the multiverse ensemble.  Its actual value in our universe is random, selected from some  distribution with a probability that is strongly modulated by anthropic selection effects. 

In these models, the Yukawa couplings $m_d^1,m_u^1,m_e^1$  in our  region of the landscape are fixed by some   symmetry. (Of course, from a  larger perspective, this symmetry is also  selected for from a larger number of possibilities, but it may not densely scan values of couplings.) Indeed of the few Standard Model parameters with strong ``selection pressure'' on their values\cite{Hogan:1999wh,Cahn:1996ag},   only $w$ scans. 
   The statistical program of comparing these models with data thus depends on  how  the   world   changes with $w$.  The ``friendliness'' of our local landscape provides a rationale for focusing on variations only on the one parameter, $w$.

Broadly speaking, the requirement that physics allows stable combinations of particles preserving structural complexity, such as atoms in living molecules,  has been referred to\cite{Arkani-Hamed:2005yv} as the  ``Atomic Principle''. 
Chemistry is full of examples of  apparent structural miracles not derived from selection; the remarkable structures of water or of DNA, for example, really do derive from symmetry and not from tuning $w$ or any other adjustable parameter. (Those structures derive just from the Schrodinger equation, and   stay   almost the same, with a change in overall scale, even if the mass of the electron  changes by  a factor of order unity; they only change qualitatively if $w$ changes by a very large factor.) Similarly, QCD predicts stable nucleonic particles like protons and neutrons without any sharp dependence on  Standard Model parameters. These properties of the Standard Model lagrangian are interpreted as  signs that we live in a friendly part of the landscape.  
A comprehensive survey of selection factors in a general landscape would include the correlations and degeneracies with other potentially variable parameters, including cosmological parameters, in radically different worlds\cite{Aguirre:2001zx}. By taking refuge in the friendly landscape, these other variations can be frozen out.

Much of the interesting structural complexity of our world derives from the availability as as well as the possibility of  building blocks, that is, atoms, able to assemble into large nontrivial combinations. The greatest sensitivity of complexity-capable atomic composition to $w$ seems to lie in  the nuclear astrophysics  of the big bang and stars that ultimately shape the conversion of baryon number into atomic nuclei.  In this paper a few examples are chosen  where there   are  unusually sudden, large qualitative responses in world behavior to small changes in $w$.

Most consequences of changing $w$,   such as modulation of weak interaction rates, change gradually, as a power of $w$.   However, the examples here display two ways in which small changes in $w$ lead to abrupt changes in overall    behavior.  One type of situation involves  discontinuous thresholds, 
 especially concerning the stability and reactions of nucleons,  where the energy balance of key reaction pathways changes suddenly;
thus it is of interest to estimate how close  the world may be  to some major thresholds of qualitatively different behavior as $w$ varies. Another type of situation involves delicately balanced astrophysical systems where an important   outcome depends on a power of $w-1$ rather than of $w$;  the  example discussed below is the famous triple-alpha tuning.

 This paper addresses  both quantitative estimates of the thresholds themselves (e.g., how big a change in $w$   destabilizes the proton?), and of the way the world would be once those thresholds are crossed (e.g., how does a neutron-stable world behave?), 
 as the masses are tuned via variations in $w$.  
 The changes in $w$ contemplated here--- which lead to altogether qualitatively different worlds--- are larger than those studied in the context of observational constraints  on time or space variations in $w$ or other Standard Model parameters in our universe (e.g., \cite{Dixit:1987at,Kujat:1999rk,Dent:2001ga,Ichikawa:2002bt,Uzan:2002vq,Yoo:2002vw,Kneller:2003ka,Dmitriev:2003qq,Langacker:2003mt,Muller:2004gu,Gassner:2006fr}.)  In addition this paper is does not actually apply anthropic reasoning or thoroughly explore the parameter space of possible worlds  (e.g.,\cite{Aguirre:2005cj}); it just explores the nuclear astrophysics of varying $w$ with other parameters held fixed. The  limited scope of the variation considered here may be too conservative for use as a real anthropic constraint in many models\cite{Harnik:2006vj}. Even so, these arguments can be invoked in the opposite sense:  the conclusion is that extremely fine tuning is not needed even for some effects where it  seemed to be.

  The arguments  here  add some new effects to    earlier studies of $w$ variation.\cite{Agrawal:1997gf,Agrawal:1998xa,Jeltema:1999na}
  The dependence of the neutron-proton mass difference on $w$  is derived, based on recently improved QCD calculations\cite{Beane:2006fk}, allowing an estimate of the threshold at $w<1$ where the neutron and not the proton becomes the stable baryon.   Astrophysical nucleosynthesis  in the neutron-stable world is shown to  lead to baryons mostly in the form of  heavy, often radioactive $r$-process and $s$-process nuclei, and not in normal stars and planets. These changes are likely to have  radical consequences for molecular evolution.

A brief survey is also presented of some  astrophysical thresholds  at   $w>1$. As $w$ increases, the neutron, and nuclei containing neutrons, become less stable. The gentlest threshold is crossed when $pp$ and then $pep$ reactions become endothermic, and deuterium is unstable to fission; next comes the point where deuterium is not even metastable long enough to function in a reaction network.  Each of these thresholds creates qualitative differences from our world, although they   may not have a significant impact on selection.

Finally,   a simple model of   resonances in $^{8}{\rm Be}$ and  $^{12}{\rm C}$, based on   nonrelativistic oscillators consisting of interacting $\alpha$ particles, and of  the Salpeter-Hoyle nucleosynthesis process where those resonances play a critical role,  is used to estimate  the $w$-dependence of the relative abundance of C and O nuclei produced in quasi-equilibrium stars. 
An  argument tracing the nuclear physics back to Standard Model parameters suggests that stellar nucleosynthesis of carbon changes significantly if $w$ increases by more than a  few percent (but not if it changes by  much less than this, as is sometimes claimed).

\section{nuclear astrophysics  with $w<1$}

\subsection{ Dependence of the n-p Mass Difference on $w$}

Many  properties of varying-$w$ worlds cannot yet be predicted very precisely, given current uncertainties in nuclear physics.  In particular, variations of $w$ lead to variations in the pion mass and therefore the range of nuclear forces. Under these circumstances, nuclear theory is not mature enough to predict how combinations  of nucleons behave and interact in detail.\cite{Beane:2002vs,Beane:2002xf}  We do not know for example at what value of $w$ (if any)  the dineutron becomes stable, or what its binding energy or decay time to a deuteron would be.\cite{Kneller:2003ka} (It is unbound in our world by only 70keV).
Even so,  the state of QCD modeling for single isolated baryons is now good enough to estimate the  neutron-proton mass difference and therefore the threshold of neutron stability.

(Uncertainties in hadron and nuclear modeling motivated some previous studies\cite{Hogan:1999wh}  to compute reaction thresholds in the plane that holds the sum of the up and down quark masses, and therefore the pion mass, constant; they were  not able to address $w$ variation, but only Yukawa variation.)

For neutrons and protons,       real {\it a priori} calculation is now possible of the mass difference in other worlds, although the errors are still large. 
The neutron-proton mass difference due to strong isospin take the form\cite{Beane:2006fk}

 \begin{equation}
 M_n-M_p|^{d-u}=(2/3)(2\bar\alpha-\bar\beta)\left({1-\eta\over1+\eta}\right)\ m_\pi^2
 \end{equation}
where the coefficients $\bar\alpha,\bar\beta$ depend only on QCD, and $\eta=m_u/m_d$ depends only on the ratio of Yukawa couplings, so the whole $w$ dependence is through $m_\pi$.
Lattice calculations with analytical extrapolation leads to the numerical estimate in our world\cite{Beane:2006fk}
\begin{equation}
M_n-M_p|^{d-u}\equiv \Delta M_{QCD} =2.26\pm 0.57\pm 0.42\pm 0.1\ {\rm MeV},
\end{equation}
where the first error is statistical, the second due to uncertainty in $\eta$, and the third due to the chiral extrapolation. 

The total mass difference also includes an electromagnetic contribution, the extra mass given the proton by its electrostatic  self-energy.   When combined with the QCD isospin breaking difference it must yield the physically measured difference, $1.2933317\pm 0.0000005$ MeV.  To the extent that the spatial distributions of quarks in nucleons is determined by their strong interactions, we expect the electromagnetic contribution to be insensitive to $w$. There is a weak dependence on $w$ because the proton is able to reduce its electromagnetic self-energy somewhat by spawning virtual $\pi^+$, and the range of this spreading depends on $m_\pi$. 
Because the pion mass is significantly larger than the electromagnetic energy term itself, we expect this contribution to scale as  $\propto (m_\pi/m_{\pi 1})^\epsilon$ where $\epsilon<<1$, and where $m_{\pi 1}$ denotes the value in our world.

The main variation of the mass difference with  $w$ is from 
 the pion mass term in the QCD contribution.  Since $m_\pi \propto \sqrt{M_p (m_u+m_d)}$, in other worlds it varies in leading order like $m_\pi=m_{\pi 1} w^{1/2}$.  Combining the QCD and electromagnetic parts, we can write for the mass difference
\begin{equation}
M_n-M_p= w\Delta M_{QCD}-w^{\epsilon/2}(\alpha_{em}/\alpha_{em1})[\Delta M_{QCD}-1.2933317\ {\rm MeV} ],
\end{equation}
where $(\alpha_{em}/\alpha_{em1})$ denotes the ratio of the electromagnetic fine structure constant to that in our universe.  (It is carried here to show the insensitivity to variations in $\alpha$ and will not be used in what follows.)

Using this, we can make an estimate of the  $w$  threshold for neutron-stable worlds. For the central value $\Delta M_{QCD}=2.26$ MeV, the threshold 
$M_n=M_p$ occurs at $w_{n=p}=0.5$; for a larger value (about $2\sigma$)  of $3.26$ MeV,  it occurs at $w_{n=p}=0.64$. 
For protons to be stable even in the presence of electrons, as in isolated atoms or neutral plasmas, we     require $M_n>M_p+m_e$ to avoid a neutron world catastrophe. This adds a term $w\times 0.511$ MeV to the threshold energy criterion; it changes the threshold estimates to $w_{n=p}=0.64$ and $w_{n=p}=0.78$ respectively.
Within the current errors of the QCD calculation, it is still possible that the threshold is significantly smaller than $w_{n=p}=0.5$. Taken together, we quote $w_{n=p}\approx 0.6\pm 0.2$ as a range of typical values.

\subsection{Neutron-stable worlds}
Important features of cosmic nuclear evolution change qualitatively as $w$ is reduced below unity. For example, big bang nucleosynthesis rather quickly becomes dominated by helium instead of hydrogen.\cite{Yoo:2002vw}  
In their survey of anthropic constraints on $w$, ABDS\cite{Agrawal:1997gf}
  concluded that anthropic lower bounds on $w$ are weak, since heavy elements are stable even for $w<<1$: even if protons in isolation are unstable, they are   stabilized within heavier nuclei, so  molecular life is   possible  in these universes. (Molecules can even contain hydrogen, in the form of deuterium, so long as the deuteron is stable.)   
However,  changes in  the overall behavior of baryonic matter  with $m_n<m_p$  would radically change cosmic evolution in other ways that have not been discussed.

 Neutrons move and carry baryon number relatively freely  through radiation and charged-particle plasmas; there is no long-range Coulomb force to inhibit nuclear interactions of neutrons with each other or with other nuclei, at virtually any temperature or density; and because neutrons can be embedded within atoms,  there is no atomic-scale limit to the density of  stable cold states  of matter, so neutron matter tends to settle into metastable systems right up to nuclear density. (They are still only metastable, because nuclear reactions do slowly create heavy charged nuclei.) Ubiquitous types of quasi-equilibria in macroscopic astrophysical objects, in particular main-sequence stars, do not  in general exist in neutron worlds as they do in our world where the baryonic stable particles are all electrically charged.

First of all, consider Big Bang nucleosynthesis in a neutron-stable world.  The main features of what happens can be found in the  AHS\cite{AHS88} study of neutron-dominated nucleosynthesis.  Bear in mind however that their model was in somewhat different context:  a neutron-rich  pocket of a (nonstandard) inhomogeneous Big Bang, but with normal  physics, including an unstable neutron. (The neutron-rich environment in their model was caused by neutron diffusion; over a wide range of scales,   neutrons but not protons diffuse before or during nucleosynthesis to populate nonlinear voids in the baryon distribution.)  

At weak decoupling (temperature $T\approx 1$ MeV, time $t\approx 1$ sec; the value of this scales only slowly with $w$), there is an excess of neutrons over protons, by an amount depending on the mass difference between the two. Additional free protons gradually continue to decay into neutrons, so the bulk of baryons become neutrons.    Residual protons become stabilized only when they can bind  into stable deuterons, which in turn consume further neutrons along a neutron-rich pathway of reactions. (The temperature at which this starts, $T\approx 0.1$ MeV, $t\approx 100$ sec in our world,  is set by thermal dissociation and is  proportional to the deuteron binding energy.)  
The main reaction path at the beginning of the process is:
\begin{equation}
p(n,\gamma){\rm D}(n,\gamma)^3{\rm H}(d,n)^4{\rm He}(t,\gamma)^7{\rm Li}(n,\gamma)^8{\rm Li}(\alpha,n)^{11}{\rm B}(n,\gamma)^{12}{\rm B}(\beta)^{12}{\rm C}(n,\gamma)^{13}{\rm C}(n,\gamma)^{14}{\rm C}
\end{equation}
The main reaction pathway might change in a modified world, but it seems likely that the main behavior would not change:
for low mass nuclei  the neutron capture cross sections  are rather small, leading to slow growth of nuclei at first, growing more rapid as the neutron capture rate increases with more massive nuclei.  

Because there is no Coulomb barrier, these reactions can continue to heavy nuclei in spite of the relatively low temperature and baryon density.  Indeed, the neutron-rich environment and long duration make this system a good site for copious production of $r$-process elements. The neutron-stable world is an even more favorable environment for this than the AHS domains with standard physics. The neutrons do not decay so there is a longer time  for them to be captured.

AHS  described  the Big Bang  $r$-process as a ``match-fuse-bomb'' sequence.  
The match is the initial growth by neutron addition to isotopes of neon, a series of slow reactions with a   small fractional yield; the fuse adds neutrons at an increasingly rapid rate until the neon seeds reach a threshold in the $A\approx 60$ region; the bomb is an exponential absorption of remaining neutrons by massive nuclei undergoing fission cycling.

 As in all $r$-process sites, the growth is regulated by a combination of neutron availability and beta-decay, but the overall behavior is dominated by the exponential at the end.  If the fuse never reaches the bomb, the bulk of the neutrons remain as free neutrons. If it does, the bulk of the neutrons remain free until the  final exponential fission cycling process, whereby  heavy nuclei (with cross sections of order barns, hundreds of times larger than the light nuclei)  capture neutrons, double their mass and fission, on a timescale of order seconds. Each fission produces two new seeds which can start the rapid capture again. Because the heavy nuclei multiply exponentially with a rate much faster than the expansion rate, it is plausible that the bulk of the baryons   end up in the very heavy elements of the massive $r$-process peaks.

The interaction time for a neutron with species $X$ in the primordial plasma, at the temperature of the cosmic background radiation, assuming thermal particle velocity $v$,  is about
\begin{equation}
t_{nX}=(n\sigma v)^{-1}\approx 5\times 10^{15}\ {\rm sec}\  (n_X/n_b)^{-1}\sigma_{X24}^{-1}z_{1000}^{-7/2},
\end{equation}
 where $n_X$ and $n_b$ denote the mean densities of species $X$ and cosmic baryons, $\sigma_{X24}$ denotes the process cross section in units of $10^{-24}{\rm cm}^2$, and $z_{1000}=(1+z)/1000$ denotes the inverse cosmic scale factor or redshift. The ratio of the interaction time to the expansion time $H^{-1}$ is about
 \begin{equation}
Ht_{nX}\approx 10^{2}\  (n_X/n_b)^{-1}\sigma_{X24}^{-1}z_{1000}^{-3/2}.
\end{equation}

Thus the system tends towards one of two behaviors. If $(n_X/n_b)$  for species with a capture cross section $\sigma_{X24}\simeq 1$  crosses a critical, exponentially small threshold allowing many fission cycles,  the baryons almost all end up in massive fission-cycled species long before reactions become slower than $H$. If not, 
 that is,  if neutrons  escape exponential capture into massive nuclei in the Big Bang, the reactions are slow enough that baryons tend to remain as mostly free neutrons, mixed with some D and He. 
 
Within the uncertainties of nuclear physics in the neutron world, it may be that the ``fuse'' stage of the $r$-process takes a long time compared to an  expansion time, so the Big Bang itself  is not efficient at consuming neutrons into heavy elements. The calculations of AHS suggest  that a small fraction of carbon and other life-sustaining elements might be produced, but in this situation the bulk of of the baryons are still neutrons as they emerge from the Big Bang.  

Since neutrons are stable, there is no time limit to their opportunity to accrete into nuclei. The generic behavior is not sensitive to  the details of the nuclear physics: the neutrons, and therefore the bulk of the baryons,  tend to agglomerate preferentially onto those nuclei with the largest cross sections, which tend to be  the large, loosely bound, massive species. This is true whether or not those species are actually stable; if they are not, it simply leads to exponential  fission cycling. It might not happen fast enough to be called an r-process  (indeed, at late times it  more closely resembles an $s$-process since it occurs closer to the valley of beta stability) but the net result is still a preponderance of heavy elements.

If the primordial $r$-process does not consume all the neutrons, the charged species D and He follow familiar behavior first as primordial ionized plasma, and eventually gravitational collapse into galaxies. The neutrons on the other hand tend to segregate from  the D and He.  
 They decouple from the plasma both thermally and kinematically before recombination,  and move for a time just  under the influence of gravity, like dark matter. However, when the plasma collapses into galaxies, the column density of typical systems is big enough to entrain many neutrons in plasma as it collapses. Those systems then become sites of further neutron capture reactions.
 
 Since nuclear energy release by neutron capture does not require high temperature or density, the energy released by decays and fission, of the order of MeV per nucleon over time, tends  to keep baryonic material hot and diffuse, and suppresses formation of stars and planets.
 In our world, planets stabilize because of electron Fermi pressure of cold atoms. In the neutron-stable world, systems start to release nuclear energy when still diffuse, delaying further collapse to planets in a dense cold state.
Stars do not collapse to a high enough density or central temperature for nuclear fusion to occur, and when they do finally reach high density, their composition is heavy nuclei whose fusion cannot support enough energy production for a main-sequence-like phase of evolution.  Smaller gravitationally bound systems  do not achieve  a cold solid or liquid state (like planets)  until nuclear neutron-capture energy sources are exhausted over time. This whole scenario is very different from our world.
 
A rough estimate shows that   even at the very low mean density of galaxies, there are already more than enough reactions with neutrons to suppress free-fall collapse and fragmentation.  The cross section for entrainment, nuclear reactions and photon scattering by electrons are roughly the same order of magnitude, $\sigma_{24}\approx 1$.  A galaxy of ionized D+He collapses only until its   density is   large enough  that the gas reacts with enough neutrons to heat it up and stop further collapse.  For typical galaxy binding energy (which ultimately derives from the primordial fluctuation amplitude), this corresponds to about 100eV of heat per baryon per cooling time, which in turn must exceed the photon escape time. With about an MeV available per nuclear reaction, at most one in 1 MeV/100eV$\approx 10^4$ baryons   reacts per photon escape time (which in this case, unlike a normal star limited by diffusion,  is about a light crossing time.)  The reactions  stabilize the galactic-scale system against collapse or fragmentation for at least $10^4$ light crossing times. This diffuse state lasts until the neutrons are mostly consumed and the baryons convert   into heavy nuclei (for the same reason as before, that once they form they dominate the time-integrated cross section for neutron absorption). For fixed composition, the reaction time scales like $\rho^{-1}$ and the collapse time like $\rho^{-1/2}$, so gravitational collapse into smaller scale systems is suppressed as long as neutrons are available to add nuclear energy in response to density increases. 
 
The late release of nuclear energy into a neutron-stable world is  dominated not by normal nuclear fusion of light elements in stars, but by neutron absorption,  beta decay and fission of massive nuclei, which ultimately comprise the bulk of baryonic matter. Eventually, as the neutrons are consumed,   baryons can settle into lower-entropy, self-gravitating systems dominated by heavy nuclei. However, that   happens only as the free energy for nuclear reactions is exhausted. Since low entropy systems in this world are still subject to the Chandrasekhar limit of only $1.4$ solar masses, but are collapsing from galactic-scale clouds without nuclear energy sources, a likely fate for much of the material is    collapse to massive black holes, at least until there is enough feedback to slow that process down with gravitational energy release. 
(This also would occur in the first case, where the Big Bang already burned everything to heavy $r$-process elements, exhausting nuclear fuel sources.)
  There may at the end be some residue of cold planetary-scale bodies made mostly of $r$-process and $s$-process elements, but   in any case there is never an opportunity for the equivalent of main-sequence stars to form.

The neutron-stable world, while very different from ours, still displays  complex astrophysical behavior on a macroscopic scale and possibly a diverse chemical composition.
Such worlds are not necessarily sterile, but it is far from clear whether combinatorial complexity comparable to our world develops at the molecular level. The transition to neutron-stable behavior abruptly raises enough difficulties for living molecules that it makes sense to adopt it as a  reliably calculable threshold for the purposes of imposing an anthropic bound, in spite of  the possible presence  of a certain amount of normal elements such as carbon and oxygen. Apart from the radical structural and chemical differences from our own world,  the ubiquitous destructive effects of neutron-activated radioactivity add additional obstacles to preservation of  DNA or RNA sequences, or indeed any other form of molecular information.  

\section{ Deuteron Instability at  $w>1$: $pp$ and  $pep$ reactions}

At  $w>1$,  neutrons start to be less stable even within nuclei. As the quark masses increase, the energy penalty for baryon number to exist as neutrons eventually exceeds the binding energy of nuclei. Deuterium is a special case; it is so loosely bound that it is affected significantly  by even small modifications in nuclear potential. 
It is also a step in the critical reactions in the early steps of nuclear burning that convert primordial hydrogen into heavier nuclei, the most important nuclear energy generation processes in stars. Once again, we omit discussion of $w$ dependence via changes in rates and cross sections that change gradually with $w$, and look instead at the thresholds where qualititative shifts may in principle be quite sudden.

Estimates of these thresholds in $w$, because they depend on interactions between nucleons,  cannot yet be computed from first principles as in the previous discussion.  In addition, although specific consequences  for nuclear reaction chains can be identified, the overall  effects on cosmic evolution are much less clear, due to uncertainties in stellar evolution models.  
Thus it is  not proven that these thresholds  have major anthropic consequences.
 It is useful nevertheless to summarize rough estimates of the sensitivity of these processes to $w$.

Tracing back to Standard Model parameters for these reactions is less reliable than for the above derivation of $w(M_n=M_p)$, because equivalent QCD estimates have not been made for the bound nucleon states. Let  $m_u^1$ and $m_d^1$  denote the  values of the quark masses, at the energies prevailing within nucleons, in our $w=1$ world. (These quantities are then proportional to the Yukawa couplings, which are here assumed to be fixed in the other worlds.)  
Then $m_u=wm_u^1$ etc.; with different $w$ we adopt a model where the proton and neutron masses change as:
\begin{eqnarray}\label{eqn: masses}
M_p&=&M_p^1-2m_u^1-m_d^1+w(2m_u^1+m_d^1)\;,\\
M_n&=&M_n^1-2m_d^1-m_u^1+w(2m_d^1+m_u^1)\;,\\
m_e&=&wm_e^1.
\end{eqnarray}
The equations (7-9) are regarded here as defining the masses $m_u^1,m_d^1$, in terms of the response of the neutron and proton masses to changes in Yukawa couplings and to $w$. The previous calculation suggests that $m_d^1-m_u^1\approx 2.6$ MeV.

The deuteron's change in mass can be decomposed into the change in masses of its constituents, and its binding energy:
\begin{equation}
M_D-M_D^1=(M_p-M_p^1)+(M_n-M_n^1)-(B_D-B_D^1)
\end{equation}
where $B_D^1=2.224644\pm0.0000334$ MeV.
Stability of the deuteron to beta decays requires
\begin{equation}
M_D<2M_p+m_e+m_{\bar\nu};
\end{equation}
in the following we will for convenience ignore the neutrino contribution.

The binding energy changes to first order in $w-1$ can  be written
\begin{equation}
B_D\approx  B_D^1 - a(w-1).
\end{equation}
The parameter $a$ is highly uncertain.
ABDS\cite{Agrawal:1997gf} presented some  simple models allowing estimates of $B_D-B_D^1$, based on first order dependence from the change in mass of pion and other mesons, and their effect on the range of force binding $D$.  They presented
numerical estimates
$a\approx 5.5 {\rm MeV}$  to as low as 1.3   MeV. On the other hand 
an analysis\cite{Beane:2002vs,Beane:2002xf}  based on 
effective field theory\cite{Kaplan:1998tg} finds that within current uncertainties on nuclear structure, an even wider range of values is allowed. 

From the above inequality,
\begin{equation}
(M_D^1-2M_p^1)+(w-1)[m_d^1-m_u^1+a]-wm_e^1<0
\end{equation}
and hence, solving for $w$,
\begin{equation}
w<{m_d^1-m_u^1+a-M_D^1+2M_p^1\over m_d^1-m_u^1+a-m_e^1}
={m_d^1-m_u^1+[a+0.931 {\rm MeV}]\over m_d^1-m_u^1+[a-0.511 {\rm MeV }]}.
\end{equation}
The stability criterion (14) is the same as the criterion for the $pep$ reaction, $p+p+e^-\rightarrow d+ \nu$, to be exothermic. A stronger condition is that the $pp$ reaction, $p+p\rightarrow d+ e^+ +\nu$, be exothermic; for this a similar derivation leads to:
\begin{equation}
w<{m_d^1-m_u^1+a-M_D^1+2M_p^1\over m_d^1-m_u^1+a+m_e^1}
={m_d^1-m_u^1+[a+0.931 {\rm MeV}]\over m_d^1-m_u^1+[a+0.511 {\rm MeV }]}.
\end{equation}

The most conspicuous astrophysical effect of these instabilities is not  the long-term decay of  free deuterium, but the effect on nuclear and stellar evolution.
The instabilities involve  a weak interaction so the lifetime of deuterium is long on a microscopic timescale and even if unstable can participate in nuclear reaction networks. On the other hand, the networks change once these thresholds are crossed.

The $pp$ reaction is the first reaction in the main path for energy generation and neutron formation in stars like our Sun,\cite{salpeterpp52} and  becomes energetically disfavored at the threshold (15).  For  $a=1$ and $m_d^1-m_u^1=2.6$, the upper limit is $w<1.1$, while for $a=6$ and $m_d^1-m_u^1=2.6$,  it is $w<1.05$. This is about  as fine as the apparent tuning of this threshold at the few percent level found for the Yukawa couplings (which was not subject to the same nuclear-physics uncertainties).\cite{Hogan:1999wh}

While it might seem that turning off the main source of energy in the Sun would have important anthropic effects, stars in worlds beyond this threshold would adapt to this situation by collapsing their cores to the point where the  $pep$ reactions  turn on and provide support. The low mass threshold for hydrogen-burning   main sequence stars increases, from its value of about 0.08 solar masses in our world. The cores of higher mass stars, of the order of one solar mass, become only slightly hotter and denser, since reaction rates depend steeply on temperature. 

Beyond the (less fine tuned) $pep$  threshold (14), deuterium is still metastable, that is, $np$ reactions are still exothermic, even if $pep$ is not. Thus, in the Big Bang, where there is a relatively plentiful supply of free neutrons (even though it is reduced from standard cosmology by thermal suppression due to the larger relative neutron mass), deuterium is still long-lived enough to act as a reaction step in the chain to making helium. (For this to be the case only requires a lifetime of order the reaction time, which is of order milliseconds.)  
As long as the helium forms and survives from the Big Bang, it is still possible to synthesize heavier nuclei in stars via other reactions. 
Above the $pep$ threshold, sufficiently massive  stars find hot  enough central densities and temperatures to generate carbon via $3\alpha\rightarrow$C, which   then allows H burning via the CNO reaction chain. 
This requires a still higher mass threshold to reach the necessary temperature and density, and cores become still  hotter and denser, but  it seems inevitable that stellar nucleosynthesis occurs even in worlds beyond the $pp$ and $pep$ thresholds. Since CNO is already not negligible in our Sun, stars of solar mass are able to find a quasi-equilibrium only slightly hotter than the Sun has.

It is not clear in the modified higher-$w$ worlds how much of  the nuclear reaction products are ejected from stars since the end stages, in particular the detailed and delicate evolution paths that lead to ejection by winds, novae, and supernovae, are altered. Certainly, elements are synthesized in deeper gravitational potentials than in our universe, and there is less nuclear energy to eject them. The catastrophic nuclear detonation that powers Type Ia supernovae, and the core collapse that powers Type II supernovae, depend on details of nuclear composition of stellar cores; too much heavy material surrounding a stellar core can effectively damp a supernova.  The initial conditions of mass and composition for the remnants controlling the behavior of  both types of supernovae might be significantly affected by changes caused by crossing the  reaction network thresholds just discussed.
 Since all the stars of a given mass are hotter and denser in their cores, it could be be that a larger percentage of their products tend to  end up in black hole  remnants instead of being ejected, like very massive stars today, or in degenerate dwarfs. Overall,  the anthropic implications of these thresholds could be significant but are not quantitatively known; they may not be significant at all.
 
More severe  consequences arise if  D is not even metastable, and  does not need a weak interaction to decay. This happens if
\begin{equation}
M_D>M_n+M_p.
\end{equation}
In this case, deuterons do not even contribute significantly to a reaction path in the Big Bang so there is no source of primordial helium. The neutrons from the Big Bang are not stabilized by being parked in helium, so they decay long before stars form.  Once stars do form, nucleosynthesis requires very high central densities  that allow weak 3-body reactions with just protons. 

Metastability of the deuteron requires
\begin{equation}
M_D^1+(M_p-M_p^1)+(M_n-M_n^1)-(B_D-B_D^1)<M_n+M_p
\end{equation}
or
\begin{equation}
w-1<
{M_D^1 -M_p^1 -M_n^1\over a}={2.2215{\rm MeV}\over a}.
\end{equation}
Not surprisingly, this limit does not depend explicitly on the quark masses; no flavor changes occur in the reaction.

As $w$ increases further,  nuclei other than deuterium gradually lose their stability. Eventually, there are no stable states of baryons other than the proton. Since hydrogen does not on its own form stable molecules heavier than H$_2$,  this is a critical threshold for molecule-based life.
As
ABDS concluded, these thresholds do not appear to be at all fine tuned in $w$; it appears that $w$ could increase by a large factor without eliminating the stable valley in the chart of the nuclides. 

\subsection{Reduced Carbon Production in Normal Stars for $w>1$:  Tuning of Resonances}
In the early 1950's, Salpeter showed\cite{Salpeter1952} that a  resonant metastable state of $^8{\rm Be}$ enables red giant stars to burn helium to carbon at rather low central temperatures, $2\times 10^8$K. Hoyle followed this with the insight\cite{Hoyle1954} that a new, previously unknown resonance in $^{12}{\rm C}$ would further increase the amount of carbon from  $^8{\rm Be} +\alpha$ reactions,  and in particular, allow carbon to be produced at the still lower temperature $\approx 10^8$K estimated for red giant stars. By increasing the ratio of carbon production to carbon burning, it also leads to comparable carbon and oxygen production in this environment, close to the actual cosmic abundance ratio. Acting on Hoyle's tip,  the resonance was quickly confirmed experimentally in Fowler's laboratory at Caltech.\cite{Dunbar1953}  This episode marked an important milestone in the early development of nuclear astrophysics.\cite{fowler1984} Since carbon and oxygen are so important to living molecules, this system is often cited as an example of anthropic fine tuning. 
 
Hoyle predicted the existence of the resonance after realizing the need for it in nucleosynthesis,
 so it is easy to see why it seemed almost miraculous when it was actually discovered experimentally; it was one of the first instances of a new piece of physics emerging from the ``cosmic laboratory''.  However, it was only  later that Hoyle floated the anthropic interpretation of the effect. (The paper often cited in the anthropic literature as the primary source, Hoyle et al., Phys. Rev. 92, 1095 (1953), is simply a brief paragraph reporting that the calculated yields fit the He:C:O abundance ratio if the $^8{\rm Be} +\alpha\rightarrow^{12}{\rm C}$ resonance exists, and reporting the discovery of the resonance.) 
 
  The binding energies of the nuclei are about 100 times larger than the resonance level splittings, and  100 times smaller than the rest masses of the nuclei, so there is certainly the appearance of fine tuning.   The following brief discussion makes a rough estimate of the sensitivity of net carbon and oxygen abundances in typical stellar environments to scanning of $w$. 
 
 The metastable $^8{\rm Be}$ resonance appears $\Delta E= 91$ keV above the energy of two $\alpha$ particles at rest. The metastable $0^+$ resonance of $^{12}{\rm C}$ appears $\Delta E =288$ keV above the  energy of  the $^8{\rm Be}$ plus another $\alpha$ particle at rest. (This is sometimes referred to as the 7.644 MeV level, referenced to the zero level of ground-state $^{12}{\rm C}$.) At a  temperature of about $10^8$K, or about $10$keV, prevailing in the energy-generating regions of the $\alpha$-burning stellar cores, an exponentially small tail of $\alpha$ particles has the right energy to participate in these resonances, where they greatly enhance the reaction rates for carbon production above what would they would otherwise be without the resonances.
  
It should be recalled that the reaction rates in stars are always very small, since the fuel is consumed over a period of millions to billions of years, as needed to replace the heat lost to radiation from the surface. But without the resonances, the temperatures of the stars would have to be  a bit larger to achieve the same energy generation rate, and at the larger temperatures, the rate of carbon destruction via $^{12}{\rm C}(\alpha,\gamma)^{16}{\rm O}$ would increase so much that very little net carbon would be left unburned. It is thus also important that there is no resonance of oxygen-16 similarly placed  within a  few hundred keV above the sum of the (ground state) $^{12}{\rm C}$ and $\alpha$ masses, since that would also destroy the carbon.

 The fact that the resonances exist is not extraordinarily surprising.
As  Weinberg\cite{Weinberg:2005fh} has recently pointed out, the existence of the $^{12}{\rm C}$ resonance is a natural consequence of nuclear dynamics, reflecting in particular the tendency of low order modes of the $^{12}{\rm C}$   nuclei to behave like collective vibrations of three $\alpha$- like constituents. A similar remark can be made about the $^8{\rm Be}$ resonance, regarding it as a mode roughly resembling an oscillator with two $\alpha$ particle masses. 
 
We conjecture  that the fact that the resonances appear slightly above the sum of the rest masses of the $\alpha$ constituents in these two nuclei is also a  natural  result of a symmetry rather than a tuning.   It is a natural consequence if the nuclei in these resonant states roughly resemble nonrelativistic oscillators of $\alpha$ particles of fixed mass with small binding energies. In that case, the oscillator energy naturally lies slightly above the summed rest masses of its constituents.  In this view,  the presence of a resonance in about the right place to enhance nuclear reactions reflects quantum mechanics of the nuclei rather than a fine tuned balance of forces or masses. Note that because ground-state $^{12}{\rm C}$, unlike $^8{\rm Be}$, is significantly bound, one would not in this view expect a similarly placed resonance in $^{16}{\rm O}$.

It should be emphasized that this view is far from rigorously demonstrated as valid in QCD-based nuclear theory of these nuclei, but it is a convenient starting point for estimates of $w$ sensitivity.
This oscillator behavior only happens if there is a (shallow) potential minimum in the separation of the $\alpha$'s,  which depends on strong and  electromagnetic couplings. However, the nonrelativistic oscillator model allows us to regard  this as a feature of the local part of the friendly landscape, not due to scanning of $w$. Because of the exponential dependence of reaction rates on energy thresholds, we will not worry about the lifetimes or widths  of these resonances.

The nonrelativistic oscillator model suggests the following estimate of the sensitivity of these effects to scanning of $w$. The potential for the oscillating $\alpha$ particles has a local minimum governed by a balance between electrostatic repulsive forces, with potential energy   $V_{es}\propto +4e^2/r$ at separation $r$, and attractive nuclear forces with a Yukawa potential $V_{QCD}\propto - \alpha_Se^{-r/r_*}$. Assuming the energy of the oscillator is   
 dominated by the exponential, and $e^2$ and $\alpha_S$ fixed, the (negative) binding energy of the oscillator relative to the zero point at infinite separation scales like $\Delta E\propto 4e^2/r^*$. The range $r_*$ of the nuclear force scales like $(m_u+m_d)^{-1/2}M_p^{-1/2}\propto w^{-1/2}$ in first order. Thus, the energy level relative to that in our $w=1$ world is
 \begin{equation}
 \Delta E= \Delta E^1 w^{1/2}.
\end{equation}
 More realistically, the true dependence is some power other than $1/2$, but as seen below, this makes only a modest numerical difference in the final answer regarding tuning.
 
 To estimate  the net impact of small changes in the resonance energy  on cosmic  nucleosynthesis, we need to account for the way the stars themselves change. As previously noted,   stars tend to adjust their core temperature and density to burn at a rate  required to support the star in  approximate hydrostatic equilibrium. In this case, they     create conditions required to burn $\alpha$'s at a certain rate because that is their source of nuclear energy. Thus the net triple-$\alpha$ reaction rate in stars in modified worlds is not sensitive to changes in the resonance energy. 
 The rate of reactions is proportional to $e^{-\Delta E/kT}$, so to keep the net rate constant,  the temperature changes from its $w=1$ value, $T_1$,  to
 \begin{equation}
T \approx T_1 w^{1/2},
 \end{equation}
 or a fractional amount
 \begin{equation}
\delta T/T \approx   (w-1)/2,
 \end{equation}
 to keep the same number of reactions in the resonance channel.
 
Rates other than those regulating the net energy release scale differently with temperature, so in general reaction ratios and abundances change.   For a reaction with an effective  Boltzman factor $e^{-E_X/T}$,  the  rate (expressed as a rate of change in species production, $\dot n$),  compared  to that in our world,   is
\begin{equation}
\dot n_1/\dot n =\exp\left[{{w^{1/2}E_{X1}-E_X}\over T}\right],
\end{equation}
so in general, the reaction rate ratios (and abundances) change with $w$ if the energy scaling differs from $E_X\propto w^{1/2}$ for some reactions. In particular if $E$ is much less sensitive to $w$, such as the energy for binding 
$^{12}{\rm C}(\alpha,\gamma)^{16}{\rm O}$ is likely to be, the differential rate scales like
\begin{equation}
\delta \dot n/\dot n =(w-1)E/2 T.
\end{equation}
 Since $E/T$ is fairly large (about 30 for carbon-producing stars), we expect fractional changes  C abundance of the order of  $\approx 15(w-1)$, a significant amount  even for changes of $w$ of less than ten percent. Requiring carbon from this source places an upper bound on $w$, since   increases greater than about five percent would burn the bulk of  carbon to oxygen at its dominant production sites. (A reduction in $w$ below unity would leave those sites carbon-rich and  oxygen-poor, but presumably leave other, hotter stars still capable of reaching temperatures for oxygen production).

These assumptions for relating rates to abundances and temperatures generally accord with the experiments of\cite{livio1989}, who ran stellar evolution codes with various  choices of $\Delta E$. This discussion has assumed local quasi-equilibrium conditions, which applies to the dominant carbon production sites in stars. Since some nucleosynthesis takes place in other very different stellar  environments (such as detonation, convection,  and/or degenerate conditions), the overall change of yield from entire  stellar populations is not clear, especially when the changes become large.
Some hypothetical changes in nuclear physics  were also shown to lead to very different carbon-to-oxygen ratio for  fractional parameter changes in nuclear potential as small as $0.4\%$\cite{Oberhummer:1999ab,Csoto:2000iw,Oberhummer:2000mn}. There is no clear contradiction with the present result since those parameters were not directly related to Standard Model parameters such as $w$.

Thus, it appears that  the production of C and O at their dominant production sites changes significantly in response to changes in $w$ at a level of about five percent, but probably not very much for changes of  less than one percent.  This remains the best candidate for an anthropically fine tuned upper limit on $w$, but it is still not as fine as sometimes claimed, and it is not airtight (since we do not really know what other sources of these elements might come into play).

\section{conclusions}
 The main points of value added here are:

1. The  dependence of the neutron-proton mass difference on $w$  is quantitatively estimated (Eq. 3).

2. Near the threshold $w_{n=p}\approx 0.6\pm 0.2$, below which the neutron becomes stable, a major qualitative shift in cosmic evolution occurs. Baryons end up predominantly in the form of very heavy nuclei, and do not form galaxies, stars and planets as we know them.

3. Thresholds in important astrophysical reactions such as $pp$ and $pep$ are crossed with $w$ increased  on the order of ten percent; but eliminating these reactions might not radically change habitability when overall stellar population behavior is considered.

4. Carbon produced in normal stars  changes  by a fraction about $15(1-w)$. In the absence of an identified compensating factor, increases in $w$ more than a few percent lead to major changes in overall cosmic carbon creation and distribution.

Within the specific context of a landscape where only $w$ varies,   worlds with nuclear astrophysics qualitatively similar to ours are expected to occur if $w$ is scanned in a relatively flat distribution, with sampling as fine as a few percent.  No other phenomenon has yet been identified requiring  ``atomic principle'' tuning finer than this. 
 	
\begin{acknowledgements}
This work was supported by NSF grant AST-0098557 at the University of
Washington. I am grateful to  M. Rees and Trinity College, Cambridge, for inspiration and hospitality, and to  E.  Salpeter, M. Savage,  and S. Weinberg for stimulating correspondence and comments.

\end{acknowledgements}

\end{document}